# New relations between analyticity, Regge trajectories, Veneziano amplitude, and Moebius transformations

Abdur Rahim Choudhary

*Bell Laboratories, 6426 Grendel Place, Bowie, Maryland 20720, USA*

In this paper we use the analyticity properties of the scattering amplitude $f(s,u)$ in the context of the conformal mapping techniques. The Schwarz-Christoffel and Riemann-Schwarz functions are used to map the upper half $s$-plane onto a triangle. We use the known asymptotic and threshold behaviors of the scattering amplitude to establish a connection between the values of the Regge trajectory functions and the angles of the triangle. This geometrical interpretation allows a link between values of the Regge trajectory functions and the generators of the invariance group of Moebius transformations associated with the underlying automorphic function. The formalism provides useful new relations between analyticity, geometry, Regge trajectory functions, Veneziano model, groups of Moebius transformations and automorphic functions. It is hoped that they will provide avenues for further work.

## Introduction

The superstring theory that has evolved into M-theory [1], has its genesis in a representation for the four point scattering amplitude by Veneziano [2]. The Veneziano formula has intrigued the physicists ever since, and a considerable effort has been spent in working backward to understand the underlying principles and symmetries [3, 4] (also see references 1 and 7).

The basic physics for our formalism is provided by the analyticity properties of the amplitude $f(s,u)$ in the s-plane, and its threshold and asymptotic behaviors. We employ the conformal mapping techniques [5, 6] to transform the upper half s-plane to a triangle in an $\omega$-plane. The transformation function $\omega$ depends on the variable s and the angles of the triangle. We use arguments based on the behavior of the amplitude at the threshold and asymptotic energies to relate these angles to the values of the Regge trajectory functions. The result is a coherent geometrical picture that throws new light on the Regge trajectory functions, Veneziano model, the ghost elimination condition, and the Pomeranchuk trajectory. In addition, the formalism has rich mathematical connections to the discontinuous groups of Moebius transformations, fundamental regions, and automorphic functions. We hope that this will provide new incentives to the string and membrane theory efforts [7] (see also chapter 2 in reference 1).

## Analyticity and conformal mapping

Consider the four point scattering amplitude $f(s,u)$ as a function of the Mandelstam variables [8] s, u, and t with $s + t + u = 4m^2$, where m is the mass of each scattering particle in the equal mass case. In this treatment we will take the momentum transfer variable t to be fixed and let s be the energy variable. Such a function has well known analyticity properties as a function of its complex arguments (see reference 8). It has two branch point singularities at $s = s_0$ and $s = u_0 \leq 0$ as well as one branch point at $\infty$. These analyticity properties are shown in Figure 1a. The cut along the right hand real s-axis corresponds to the s-channel interactions, while the cut along the negative real s-axis arises from the requirement of crossing symmetry and represents the t-channel interactions. We have considered only the singularities corresponding to the two particle thresholds.

First, we map the s-plane onto a z-plane such that the three branch points $s_0$, $u_0$ and $\infty$ are respectively mapped onto 1, 0, and $\infty$.

$$z = \frac{s - u_0}{s_0 - u_0} \quad (1)$$

The cut z-plane is shown in figure 1b.

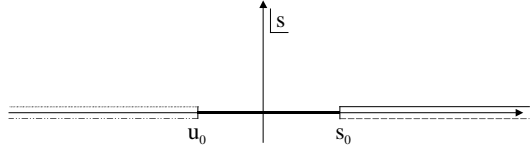

Figure 1a

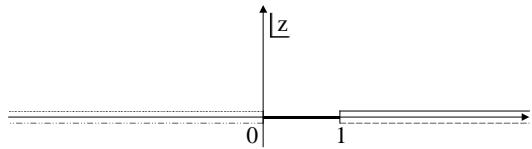

Figure 1b

FIG. 1. The analyticity of scattering amplitude in the (a) s-plane (b) z-plane

Next we map the upper half cut z-plane ($\operatorname{Im} z > 0$) onto the triangle DEF in $\omega$-plane, as shown in Figure 2. The existence and construction of this mapping is proven by the Schwarz-Christoffel theorem [9].

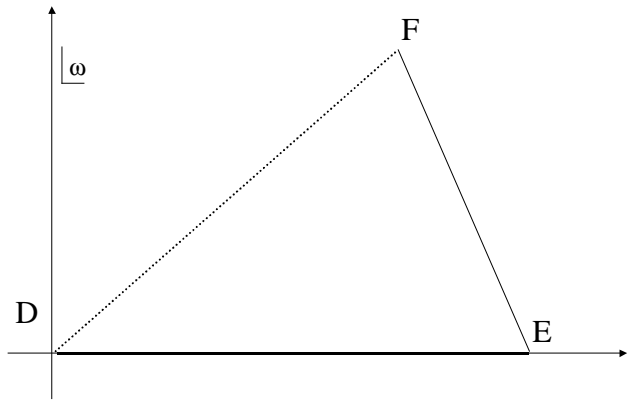

FIG. 2. The upper half z-plane is mapped on to the triangle DEF in the $\omega$-plane

The mapping function in the present case is referred to as Schwarz triangle function, and it is given by the incomplete Euler beta function as follows

$$\omega(z) = \int_0^z x^{A-1}(1-x)^{B-1} dx \qquad (2)$$

The transformation represented by equation 2 maps the upper half z-plane onto the triangle DEF such that the branch points at 0, 1, and ∞ in the z-plane are respectively mapped onto the vertices D, E, and F of the triangle DEF in the $\omega$-plane.

The parameters A and B in equation 2 are related to the interior angles of the triangle DEF as follows.

$$\pi A = \angle EDF \qquad (3a)$$

$$\pi B = \angle DEF \qquad (3b)$$

They determine the precise behavior of the amplitude at its branch points.

We now introduced a quantity C as follows.

$$\pi C = \angle DFE \qquad (3c)$$

Obviously we have the following Euclidean relation.

$$A + B + C = 1 \qquad (4)$$

### Relation to Regge Trajectories

In order to determine the transformation function $\omega$ represented by equation 2, we need to determine two of the three angles of the triangle DEF. We will do this by requiring that the transformation function $\omega$ possess the behavior of the amplitude at the threshold and asymptotic energies. Since the amplitude $f(s,u)$ would eventually be described as a function of $\omega$, it makes sense to build as much of its analyticity as possible into the transformation variable $\omega$. As discussed in reference 6, this in fact is the main idea behind the multi-Riemann-sheet conformal mapping techniques.

For this purpose, let us examine the behavior of the conformal mapping variable $\omega$ at the vertices D, E, and F which respectively correspond to the two particle threshold in the t-channel, the two particle threshold in the s-channel, and the asymptotic energies in the s-channel.

### Asymptotic behavior

Let us consider the behavior of equation 2 for asymptotic values of energy s. The point s=∞ is mapped onto z=∞, which is mapped onto the vertex F in the ω-plane. In the neighborhood of z=∞, the mapping function ω behaves as follows

$$\omega(z) = \omega(\infty) + [\frac{1}{z}h(\frac{1}{z})]^{C-1} \qquad (5)$$

where C is given by equation 3c and $h(\frac{1}{z})$ is regular and non zero at $z = \infty$.

Asymptotically, the amplitude is known to have the Regge behavior [10]

$$f(s,u) \xrightarrow[s \to \infty]{} s^{\alpha_t(t)} \qquad (6)$$

where $\alpha_t(t)$ is the t channel Regge trajectory function.

Comparing equations 5 and 6 and recalling equation 1 we obtain

$$C = 1 - \alpha_t(t) \qquad (7)$$

The momentum transfer variable $t \leq 0$ is fixed. As t is fixed at different values, the geometry of the triangle in figure 2 changes accordingly. This use of a variable geometrical region for the purpose of conformal mapping techniques is one distinguishing aspect of our formalism compared to other conformal mapping methods used in particle physics [11].

### Threshold behavior

Let us next consider the behavior of equation 2 at the threshold energy. The s-channel threshold $s_0$ is mapped onto the point z=1 which in turn is mapped onto the vertex E of the triangle. In this neighborhood, the $\omega$ of equation 2 behaves as follows

$$\omega(z) = \omega(1) + [(z-1)h_1(z-1)]^{B-1} \qquad (8)$$

where $h_1(z-1)$ is a function regular and non zero at $z = 1$, and B is given by equation 3b.

The behavior of the amplitude at this point is the behavior of the Legender polynomial $(P_{\alpha_s(s)}(\cos\theta_s))$ at the s-channel threshold $s=s_0$, as it is analytically continued from the point $\alpha_s(s) = \mathrm{int}\,eger$, $\cos\theta_s = 1 + \frac{2t}{s - s_0}$ to the unphysical point $t \neq 0$, $s=s_0=4m^2$. Noting that $\cos\theta \xrightarrow{s \to s_0} \infty$ we have (see reference 8)

$$f(s,u) \xrightarrow[s \to s_0]{} (s - s_0)^{-\alpha_s(s_0)} \qquad (9)$$

where $\alpha_s(s)$ is the s channel Regge trajectory function.

Comparing equations 8 and 9 and recalling equation 1 we obtain

$$B = 1 - \alpha_s(s) \qquad (10)$$

where $\alpha_s(s)$ is evaluated at the threshold $s=s_0$ for the purpose of determining the geometry of the triangle.

## Ghost elimination condition

Let us introduce a u channel Regge trajectory function $\alpha_u(u)$ as follows

$$A = 1 - \alpha_u(u) \qquad (11)$$

where A is given by equation 3a.

Equations 7, 10, and 11 provide a geometrical interpretation for the Regge trajectories. Together with equation 4, they yield the following relation

$$\alpha_s + \alpha_t + \alpha_u = 2 \qquad (12)$$

This represents condition to eliminate ghosts from the Veneziano amplitude (see reference 2). In our treatment it arises as a geometrical constraint.

## Pomeranchuk trajectory and the threshold behavior

The threshold points $s_0$ and $u_0$ in figure 1a would become the physical two particle thresholds for the case of zero momentum transfer, t=0. The branch points at these two particle thresholds are of the square root type, as shown in reference 8. To obtain a square root behavior at $s = s_0$ we use equation 8 and 9, and require

$$1 - B = \alpha_s(s = s_0 = 4m^2) = \tfrac{1}{2} \qquad (13)$$

Similarly, for the square root branch point at $s=u_0$, we obtain the following result

$$1 - A = \alpha_u(u = u_0 = 0) = \frac{1}{2} \qquad (14)$$

Invoking equation 12 we obtain

$$\alpha_t(t = 0) = 1 \qquad (15)$$

This is a Pomeranchuk type trajectory with unit intercept. In our formalism, it arises from the square root type branch points at the two particle physical thresholds in the s and t channels. Pomeranchuk trajectory plays an important role in understanding the high energy behavior of total cross sections in the forward direction [12] (also see chapter 3 of reference 8).

Equations 13-15 show that, in the special case of t=0, the triangle in figure 2 degenerates into a vertical strip such that the angles at the vertices D and E are each a right angle and the angle at the vertex F is zero. At the same time the mapping function given by equation 2 reduces to the following trigonometric function.

$$\omega_1(z) = \int_0^z x^{\tfrac{1}{2}-1}(1-x)^{\tfrac{1}{2}-1} dx = \arccos(1 - 2z) \qquad (16)$$

This representation does indeed have the square root type branch points at z=1 and 0, while it behaves logarithmically at infinity.

### Amplitude representation

As discussed above, the conformal mapping of the upper half s-plane onto the triangle DEF in the $\omega$-plane is given by the following representation.

$$\omega(z) = \int_0^z x^{-\alpha_u}(1-x)^{-\alpha_s}\,dx \qquad (18)$$

where we have substituted equations 10 and 11 in equation 2.

The value of the conformal mapping variable $\omega$ lies in its ability to afford simple representations for the amplitude $f(s,u)$. Let us denote this representation by the following equation.

$$f(s,u) = \phi(\omega) = \sum_{n=0}^{n=\infty} a_n (\omega - \omega_c)^n \qquad (19)$$

where $\omega$ is given by equation 18, and $\omega_c$ is the point around which the expansion is sought.

As we have seen above, the variable $\omega$ correctly represents the analyticity of the amplitude, i.e. it carries the two particle thresholds for the s and t channels, as well as the threshold and asymptotic behaviors. Therefore the simplest zeroeth order representation for the amplitude is as follows

$$\phi_0(\omega) = \omega \qquad (20)$$

The next improvements can be achieved recursively by using the lower level approximation in evaluating the expansion coefficients in equation 19. Two cases may be considered for illustration.

First, let us expand about the point corresponding to the two particle threshold in the s channel, which corresponds to s=$s_0$ or z=1. Using equation 20 on the right hand side of equation 19, the constant term reads as follows.

$$\frac{\Gamma(1-\alpha_s)\Gamma(1-\alpha_u)}{\Gamma(2-\alpha_s-\alpha_u)} \qquad (21)$$

Second, let us expand about the point at infinity, which corresponds to z=$\infty$ or the vertex F in figure 2. A similar procedure yields the constant term as follows.

$$\frac{\Gamma(1-\alpha_s)\Gamma(1-\alpha_u)}{\Gamma(2-\alpha_s-\alpha_u)} + (-1)^{-\alpha_s}\frac{\Gamma(1-\alpha_s)\Gamma(1-\alpha_t)}{\Gamma(2-\alpha_s-\alpha_t)} \qquad (22)$$

where we decomposed the range of integration 0 to ∞ into ranges 0 to 1 and 1 to ∞, changed the variable as $z \to \frac{1}{r}$ in the second range, and used equation 12.

The similarity of these terms with the Veneziano model is obvious. The next order terms can also be computed. These may help restore the Unitarity property of the Veneziano amplitude, in so far as they incorporate the branch point cuts.

### Relationship to Automorphic functions

Let us take another look at figure 2. It shows how the upper half z-plane shown in figure 1b is mapped onto the $\omega$-plane. Let us now consider how the entire z-plane maps onto the $\omega$-plane. This is shown in figure 3. The entire z-plane in figure 1b is mapped on to the quadrilateral DFEG in the $\omega$ plane by the transformation in equation 2, and equivalently equation 18. The upper and lower edges of the right hand cut in figure 1 have been unfolded onto the sides EF and EG respectively, of the quadrilateral DFEG. Similarly, the upper and lower sides of the left hand cut have been unfolded on to sides DF and DG respectively.

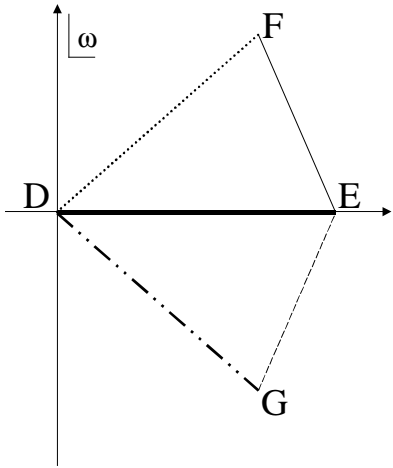

Figure 3: Mapping of the entire z-plane shown in figure 1b on to the $\omega$-plane

In the parlance of automorphic functions and discontinuous groups of Moebius transformations [13, 14], the quadrilateral DFEG is referred to as a "fundamental region". There is a group of discontinuous transformations associated with this fundamental region. The group is generated by the transformations that map the pair of conjugate sides of the fundamental region on to each other. Thus one generator of the group is the transformation that maps the side DF on the side DG. It is given by the Moebius transformation $T_1$ below.

$$T_1 = \begin{pmatrix} \exp(-i\pi\alpha_u) & 0 \\ 0 & \exp(i\pi\alpha_u) \end{pmatrix} \qquad (23)$$

The second generator corresponds to the transformation that maps the side EF on the side EG. It is given by the Moebius transformation $T_2$ below.

$$T_2 = \begin{pmatrix} \exp(i\pi\alpha_s) & (2id)\sin\pi\alpha_s \\ 0 & \exp(-i\pi\alpha_s) \end{pmatrix} \qquad (24)$$

where d is the length DE in figure 3, which is represented by the expression numbered 21 above. Thus d is given by the following equation.

$$d = \frac{\Gamma(1-\alpha_s)\Gamma(1-\alpha_u)}{\Gamma(2-\alpha_s-\alpha_u)} \qquad (25)$$

It is interesting that the values of the Regge trajectory functions enter in the generators for the underlying invariance group.

Each copy of the fundamental region DFEG under the group of Moebius transformations generated by the above generators is a complete map of the entire z-plane. The conformal mapping variable $\omega$ is thus a multivalued function of z. The inverse transformation of equation 2, or equivalently equation 18, would map the multiple values of $\omega$ on to the same value of z, and z is an automorphic function of $\omega$. The automorphic functions are a strong instrument to understand some interrelations between geometry, discrete groups of Moebius transformations, and conformal mappings (see reference 6).

At the same time, the transforms of the "fundamental region" under these transformations will completely cover the $\omega$ plane, without overlaps and without leaving any chinks (see references 13 and 14). This aspect requires that the generators $T_1$ and $T_2$ be cyclic of finite or infinite order. If a generator would have a finite order n, it implies from the structure of $T_1$ and $T_2$ that the corresponding Regge trajectory function shall take values at the threshold points $s_0$ and $u_0$ according to the following constraint.

$$\alpha = \frac{1}{n} \qquad (26)$$

### Generalization to non Euclidean geometry

The approach can be extended to the case of non Euclidean geometry, in particular the Poincare model [15], which is often used. A need to use non Euclidean geometry would be indicated if the constraint in equation 12 would read as follows.

$$\alpha_s + \alpha_t + \alpha_u > 2 \qquad (27)$$

This can correspond to the case where the generators of the underlying discontinuous group of Moebius transformations would turn out to be cyclic with infinite order. The geometry for the fundamental region would then involve curvilinear triangles. The transformation would then be represented by a general class of functions known as Riemann-Schwarz triangle functions [16]. Defining A, B, and C as in equations 3a to 3c, the relationship between the angles now reads as follows.

$$A + B + C < 1 \qquad (28)$$

The transformation from the z plane to the $\omega$ plane is now given by the following equation.

$$\omega = \tau(z) = \frac{f_1(z)}{f_2(z)} \qquad (29)$$

where

$$f_1(z) = \int_0^1 dt [ t^{-\frac{1}{2}(1+A+B+C)} (1-t)^{-\frac{1}{2}(1+A-B-C)} (1-zt)^{-\frac{1}{2}(1-A+B-C)} ] \qquad (29A)$$

$$f_2(z) = \int_0^1 dt [ t^{-\frac{1}{2}(1+A+B+C)} (1-t)^{-\frac{1}{2}(1-A-B+C)} (1-t+zt)^{-\frac{1}{2}(1-A+B-C)} ] \qquad (29B)$$

Alternative representation using the familiar Gamma functions ($\Gamma$) and Gaussian Hypergeometric functions ($F_{21}$) are as follows

$$\tau(z) = \frac{g_1(z)}{g_2(z)} \qquad (30)$$

where

$$g_1(z) = \frac{\Gamma(\frac{1}{2}(1-A+B+C))}{\Gamma(1-A)} F_{21}(\frac{1}{2}(1-A+B-C), \frac{1}{2}(1-A-B-C); 1-A; z) \qquad (30A)$$

$$g_2(z) = \frac{\Gamma(\frac{1}{2}(1+A+B-C))}{\Gamma(1-C)} F_{21}(\frac{1}{2}(1-A+B-C), \frac{1}{2}(1-A-B-C); 1-C; 1-z) \qquad (30B)$$

A special case corresponding to A=B=C=0 of this transformation was found useful in exploring the analyticity properties of the partial wave scattering amplitudes [17, 18].

## Summary and Conclusions

We have used the analyticity properties as our basic physical input to the conformal mapping techniques, determined the angles of the conformal mapping triangle in terms of the values of the Regge trajectories. We derived the conformal mapping variable and used it to arrive at the Veneziano type amplitude in the lowest order approximation, with clear paths for improvement. This provided a geometrical interpretation for the Regge trajectories, and yielded the ghost elimination condition as a geometrical constraint.

Our framework has allowed us to use the square root nature of the two particle thresholds to determine the values of the trajectory functions at these thresholds for the case of forward scattering, which led to the existence of a Pomeranchuk type trajectory with unit intercept.

Finally, a connection is made with the underlying automorphic function and the associated invariance group of Moebius transformations. We have linked the Regge trajectories to the generators of this group, which constrains the values of the Regge trajectories to take only inverse integer values at the threshold. The consequences of these new relations remain to be further explored.